\documentclass[aps, prb, twocolumn, showpacs, superscriptaddress]{revtex4-1}

\usepackage{amsmath,amssymb}
\usepackage{graphicx}% Include figure files
\usepackage{dcolumn}% Align table columns on decimal point
\usepackage{bm}% bold math
\begin{document}

\title {A note on path integral formulism of $Z_{2}$ slave-spin representation of Hubbard model}
\author{Yin Zhong}
\email{zhongy05@hotmail.com}
\affiliation{Center for Interdisciplinary Studies $\&$ Key Laboratory for
Magnetism and Magnetic Materials of the MoE, Lanzhou University, Lanzhou 730000, China}
\author{Hong-Gang Luo}
\email{luohg@lzu.edu.cn}
\affiliation{Center for Interdisciplinary Studies $\&$ Key Laboratory for
Magnetism and Magnetic Materials of the MoE, Lanzhou University, Lanzhou 730000, China}
\affiliation{Beijing Computational Science Research Center, Beijing 100084, China}

\date{\today}

\begin{abstract}
This is a note on the derivation of path integral formulism for $Z_{2}$ slave-spin representation of Hubbard model and may be helpful for further study in related works. We are rather happy to receive any comments and other discussions are welcome.
\end{abstract}

\maketitle

\section{$Z_{2}$ slave spin representation of the Hubbard model} \label{sec2}
The model we used is the Hubbard model at half-filling,
\begin{eqnarray}
&&H=-t\sum_{<ij>\sigma}(c_{i\sigma}^{\dag}c_{j\sigma}+h.c.)+\frac{U}{2}\sum_{i}(n_{i}-1)^{2}
\end{eqnarray}
where $n_{i}=\sum_{\sigma}c_{i\sigma}^{\dag}c_{i\sigma}$, $U$ is the onsite Coulomb energy between electrons on the same site and $t$ is the hopping energy between nearest sites.

In the treatment of $Z_{2}$ slave-spin approach, the physical electron $c_{\sigma}$ is fractionalized into a new auxiliary fermion $f_{\sigma}$ and a slave spin $\tau^{x}$ as\cite{deMedici,Hassan,Ruegg,Yu,Nandkishore,Zhong2012,Ruegg2012}
\begin{equation}
c_{i\sigma}=f_{i\sigma}\tau_{i}^{x}
\end{equation}
with a constraint $\tau_{i}^{z}+1=2(n_{i}-1)^{2}$ enforced in every site and $\tau^{x}$, $\tau^{z}$ are the standard Pauli matrix. Under this representation, the original Hamiltonian can be reformulated as
\begin{eqnarray}
H'=-t\sum_{<ij>\sigma}(\tau_{i}^{x}\tau_{j}^{x}f_{i\sigma}^{\dag}f_{j\sigma}+h.c.)+\frac{U}{4}\sum_{i}(\tau_{i}^{z}+1)
\end{eqnarray}
where $n_{i}=n_{i}^{f}=\sum_{\sigma}f_{i\sigma}^{\dag}f_{i\sigma}$. Obviously, a $Z_{2}$ local gauge symmetry is left in this representation (both slave-fermions and slave spins carrying the $Z_{2}$
gauge charge) and the corresponding low-energy effective theory should respect this. The mentioned gauge structure can be seen if $f_{i\sigma}^{(\dag)}\rightarrow \epsilon_{i}f_{i\sigma}^{(\dag)}$ and $\tau_{i}^{x}\rightarrow\epsilon_{i}\tau_{i}^{x}$ with $\epsilon_{i}=\pm1$ while the whole Hamiltonian $H$ is invariant under this $Z_{2}$ gauge transformation.

The aim of this note is to derive a useful path integral formulism for the above $Z_{2}$ slave-spin representation of the Hubbard model at half-filling. But before attacking this central problem, firstly, we will give a short review of the path integral treatment of the quantum Ising model in transverse field as our starting point.\cite{Stratt} Another reason of this review is that although it is well-known the quantum Ising model in transverse field can be describe by an effective $\varphi^{4}$ theory in one higher dimension, a real calculation of this statement has not been included in standard textbook on quantum phase transitions.\cite{Continentino,Sachdev2011}

\section{path integral for the quantum Ising model in transverse field}
The quantum Ising model in transverse field is defied as\cite{Sachdev2011}
\begin{equation}
\hat{H}_{I}=-J\sum_{<ij>\sigma}\tau_{i}^{z}\tau_{j}^{z}+h.c.)-K\sum_{i}\tau_{i}^{x}
\end{equation}
where a ferromagnetic coupling $J>0$ is assumed and $K$ represents the the transverse external field.

At first sight, one may directly use the coherent state of spin operator in constructing the path integral representation, (One can find a brief but useful introduction to this issue in Ref. 9.) however, this will lead to an extra topological Berry phase term and is not easy to utilize practically. An alterative approach is to use the eigenstates of spin operator $\tau^{x}$ or $\tau^{z}$ as the basis for calculation.\cite{Stratt}
One will see this approach is free of the topological Berry phase term and give rise to a rather simple formulism.  Therefore, to construct a useful path integral representation, we will follow Ref. 8.

First, we consider the orthor-normal basis of $N_{s}$-Ising spins as
\begin{equation}
|\sigma\rangle\equiv|\sigma_{1}\rangle|\sigma_{2}\rangle|\sigma_{2}\rangle\cdot\cdot\cdot|\sigma_{N}\rangle
\end{equation}
with $\sigma_{i}=\pm1$ and define
\begin{equation}
\tau_{i}^{z}|\sigma\rangle=\sigma_{i}|\sigma\rangle,
\end{equation}
\begin{equation}
\tau_{i}^{x}|\sigma\rangle
=|\sigma_{1}\rangle|\sigma_{2}\rangle|\sigma_{3}\rangle\cdot\cdot\cdot|-\sigma_{i}\rangle\cdot\cdot\cdot|\sigma_{N}\rangle.
\end{equation}
Then the partition function $Z=Tr(e^{-\beta \hat{H}})$ can be represented as
\begin{eqnarray}
Z=\sum_{\{\sigma\}=\pm1}\prod_{n=1}^{N}e^{\epsilon J\sum_{<ij>}\sigma_{i}(n)\sigma_{j}(n)}\langle\sigma(n+1)|e^{\epsilon K\sum_{i}\tau_{i}^{x}}|\sigma(n)\rangle \nonumber
\end{eqnarray}
where $\epsilon$N=$\beta$.
The calculation of $\langle\sigma(n+1)|e^{\epsilon K\sum_{i}\tau_{i}^{x}}|\sigma(n)\rangle$ is straightforward
by exponentiating the $\tau_{i}^{x}$ matrix and one gets
\begin{eqnarray}
\langle\sigma(n+1)|e^{\epsilon K\sum_{i}\tau_{i}^{x}}|\sigma(n)\rangle&&=\frac{1}{2}(e^{\epsilon K}+e^{-\epsilon K}\sigma_{i}(n)\sigma_{i}(n+1)),\nonumber \\
&&=e^{a\sigma_{i}(n)\sigma_{i}(n+1)+b}
\end{eqnarray}
where $a=\frac{1}{2}[\ln\cosh(\epsilon K)-\ln\sinh(\epsilon K)]$ and $b=\frac{1}{2}[\ln\cosh(\epsilon K)+\ln\sinh(\epsilon K)]$.
Therefore, the resulting path integral formulism for the quantum Ising model in transverse field is
\begin{equation}
Z=\sum_{\{\sigma\}=\pm1}\prod_{n=1}^{N}e^{\epsilon J\sum_{<ij>}\sigma_{i}(n)\sigma_{j}(n)+\sum_{i}a\sigma_{i}(n)\sigma_{i}(n+1)+N_{s}b}.
\end{equation}

Further, if one assumes the model is defined in a hyper-cubic lattice in space dimension of d, an effective theory can be derived as
\begin{equation}
Z=\int D\phi \delta(\phi^{2}-1)exp(-\int d\tau d^{d}x \frac{1}{2g}[(\partial_{\tau}\phi)^{2}+c^{2}(\nabla\phi)^{2}])
\end{equation}
where $\frac{1}{2g}=(\frac{a\epsilon}{a_{0}^{d}})^{\frac{d+1}{2}}$ with $a_{0}$ being the lattice constant and $c^{2}=\frac{Ja_{0}^{d-2}}{a\epsilon}$. Moreover, in the effective theory, $\phi$ corresponds to $\tau^{z}$ while $\tau^{x}$ gives the kinetic energy term in imaginary time. Then, the standard $\phi^{4}$ theory is obtain by relaxing the hard constraint $\phi^{2}=1$ while introducing a potential energy term,
\begin{equation}
Z=\int D\phi exp(-\int d\tau d^{d}x [(\partial_{\tau}\phi)^{2}+c^{2}(\nabla\phi)^{2}+r\phi^{2}+u\phi^{4}])
\end{equation}
where $r,u$ are effective parameters depending on microscopic details.

\section{path integral for the $Z_{2}$ slave-spin representation of the Hubbard model}
After reviewing the path integral representation of the quantum Ising model in transverse field, we
will turn to the construction of path integral for $Z_{2}$ slave-spin approach of the Hubbard model.
To this aim, we follow the logic of Ref. 11 where the general $Z_{2}$ gauge theory is constructed in an extended Hubbard model.\cite{Senthil}

The construction of path integral is to calculate the partition function $Z=Tr(e^{-\beta \hat{H}}\hat{P})$ where
$\hat{P}$ is the projective operator to exclude unphysical states introduced by $Z_{2}$ slave-spin representation.
Here we use
\begin{equation}
\hat{P}=\prod_{i}(1+(-1)^{\frac{1}{2}[\tau_{i}^{z}+1-2(n_{i}^{f})^{2}]}).
\end{equation}
This choice has the advantage to meet the mean-field theory of $Z_{2}$ slave-spin approach clearly though another choice can also be used equivalently
\begin{equation}
\hat{P}=\prod_{i}(1+(-1)^{\frac{1}{2}[\tau_{i}^{z}-1+2n_{i}^{f}]}).
\end{equation}
We will use the first definition of $\hat{P}$ in the following discussion.
Follow Ref. 11, the projective operator can be reformulated by introducing auxiliary Ising field $\sigma_{i}=\pm1$
\begin{equation}
\hat{P}=\prod_{i}\frac{1}{2}\sum_{\sigma_{i}=\pm1}exp(i\frac{\pi}{4}(\sigma_{i}-1)[\tau_{i}^{z}+1-2(n_{i}^{f})^{2}]).
\end{equation}
Since $[\hat{P},H']=0$, one can define a modified Hamiltonian $H_{eff}$ as
\begin{eqnarray}
H_{eff}=H'+\sum_{i}i\frac{\pi}{4}(1-\sigma_{i})[\tau_{i}^{z}+1-2(n_{i}^{f})^{2}].
\end{eqnarray}
Then using the same method in the treatment of quantum Ising model and standard coherent state representation of fermions, one obtains the path integral formulism of $Z_{2}$ slave-spin representation of Hubbard model
\begin{equation}
Z=\prod_{ni}\int d\bar{f}_{i}(n)df_{i}(n)d\varphi_{i}(n)\delta(\varphi^{2}_{i}-1)d\sigma_{i}(n)\delta(\sigma^{2}_{i}-1) e^{-S}
\end{equation}
and
\begin{eqnarray}
S&&=\sum_{ni}\bar{f}_{i\sigma}(n)(f_{i\sigma}(n)-f_{i\sigma}(n-1)) \nonumber\\
&&+\sum_{ni}\varphi_{i}(n)\varphi_{i}(n+1)a(n) \nonumber\\
&&-\epsilon t\sum_{n<ij>}(\varphi_{i}(n)\varphi_{j}(n)\bar{f}_{i\sigma}(n)f_{j\sigma}(n)+c.c.) \nonumber\\
&&+\epsilon\sum_{ni}i\frac{\pi}{4}(1-\sigma_{i}(n))[1-2(n_{i}^{f})^{2}]
\end{eqnarray}
where we use $\tau_{i}^{x}|\varphi\rangle=\varphi_{i}|\varphi\rangle$ with $\varphi=\pm1$ and $\tau_{i}^{z}|\varphi\rangle
=|\varphi_{1}\rangle|\varphi_{2}\rangle|\varphi_{3}\rangle\cdot\cdot\cdot|-\varphi_{i}\rangle\cdot\cdot\cdot|\varphi_{N}\rangle$ to avoid confusion with auxiliary Ising field $\sigma_{i}$ with $a(n)=-\frac{1}{2}\ln\epsilon(\frac{U}{4}+i\frac{\pi}{4\beta}(1-\sigma_{i}(n)))$. The above action is our main result in this note and further approximations have to be made in order to gain some physical insights. A popular approximation is to decouple the interaction term between slave-spin $\varphi$ and auxiliary fermion $f_{\sigma}$ at the mean field level and then reintroduce phase fluctuation (in fact, a $Z_{2}$ gauge field due to the gauge structure of $H'$).\cite{Senthil,Ruegg2012} However, details of this treatment has not been reported until now but we hope our formulism constructed here may be useful in this direction.

\begin{acknowledgments}
The work was supported partly by NSFC, the Program for NCET, the Fundamental Research Funds for the Central Universities and the national program for basic research of China.
\end{acknowledgments}

\end{document}